\documentclass[12pt]{article}
\begin{document}
\title{FLUCTUATIONAL SPACETIME}
\author{B.G. Sidharth\\
International Institute for Applicable Mathematics \& Information Sciences\\
Hyderabad (India) \& Udine (Italy)\\
B.M. Birla Science Centre, Adarsh Nagar, Hyderabad - 500 063 (India)}
\date{}
\maketitle
\begin{abstract}
Newton's action at a distance gravitational law and Coulomb's action at a distance electrostatic law had to be reexamined in the light of field theories which originated from Maxwell's electrodynamics. These ideas were further modified with the advent of Quantum Theory, even though the differentiable spacetime manifold of Classical Physics was retained. More recent approaches in unifying gravitation with the other fundamental interactions has lead to a spacetime that is not smooth. In this light we examine the nature of spacetime, taking into consideration the role of fluctuations.
\end{abstract}
\section{Introduction}
In Newton's conception, there was an absolute background space. While matter, forces, energy and the like were actors acting out in time. So also the law of gravitation was an action at a distance theory. Every material particle exerted the force of gravitation instantly on every other material particle.\\
In the next century, Coulomb discovered the law of electric, more precisely electrostatic interaction. It had the same form of an inverse square dependence on distance, as gravitation. This too was an action at a distance force. While the action at a distance gravitational law worked satisfactorily, in the nineteenth century the Coulomb law encountered difficulties when it was discovered by Ampere, Faraday and others that moving charges behaved differently. The stage was being set for Maxwell's electrodynamics. Maxwell could unify the experimental laws of Faraday, Ampere and others in a Field Theory \cite{jackson}. Already in the seventeenth century itself Olaf Romer had noticed that light travels with a finite speed and does not reach us instantly. He could conclude this by observing the eclipses of the satellites of Jupiter. Christian Huygens took the cue and described the motion of light in the form of waves. The analogy with ripplies moving outwards on the surface of a pool was clear.\\
Maxwell utilised these ideas in interpreting the experimentally observed laws of electricity and magnetism. Thus a moving charge would cause a ripple in an imaginary medium or field, and that ripple would be propagated further till it hit and acted upon another charge. This was a dramatic departure from the action at a distance concept because the effect of the movement of the charge would be felt at a later time by another charge. Maxwell even noticed that the speed at which these disturbances would propagate through the field was the same as that of light. Already the stage was being set for Einstein's Theory of Special Relativity \cite{einstein}. Even so, it must be mentioned that in the earlier formulation of the old action at a distance theory which resembled closely Maxwell's Field Theory, in mathematical form.\\
At this stage it was clear that two closely related concepts were important-- Locality and Causality. We will return to this shortly but broadly what is meant is that parts of the universe could be studied in isolation and further, that an event at a point $A$ cannot influence an event at a point $B$ which cannot be reached by a ray of light during this interval. Roughly speaking, all events within this light radius would be causally connected, but not so events beyond this radius.
\section{Action at a Distance Electrodynamics}
From a classical point of view a charge that is accelerating radiates energy which dampens its motion. This is given by the well known Maxwell-Lorentz equation, which in units $c = 1$, is \cite{hoyle}
\begin{equation}
m \frac{d^2x^\imath}{dx^2} = e F^{\imath k} \frac{dx^k}{dt} + \frac{4e}{3} g_{\imath k} \left(\frac{d^3x^\imath}{dx^3} \frac{dx^1}{dt} - \frac{d^3x^1}{dx^3} \frac{dx^\imath}{dt}\right) \frac{dx^k}{dt},\label{e1}
\end{equation}
The first term on the right is the usual external field while the second term is the damping field which is added ad hoc by the requirement of the energy loss due to radiation. In 1938 Dirac introduced instead of (\ref{e1}),
\begin{equation}
m \frac{d^2x^\imath}{dx^2} = e \left\{F^\imath_k + R^\imath_k\right\} \frac{dx^k}{dt}\label{e2}
\end{equation}
where
\begin{equation}
R^\imath_k \equiv \frac{1}{2} \left\{F^{\mbox{ret}\imath}_k - F^{\mbox{adv}\imath}_k\right\}\label{e3}
\end{equation}
In (\ref{e3}), $F^{\mbox{ret}}$ denotes the retarded field and $F^{\mbox{adv}}$ the advanced field. While the former is the causal field where the influence of a charge at $A$ is felt by a charge at $B$ at a distance $r$ after a time $t = \frac{r}{c}$, the latter is the advanced acausal field which acts on $A$ from a future time. 
In effect what Dirac showed was that the radiation damping term in (\ref{e1}) or (\ref{e2}) is given by (\ref{e3}) in which an antisymmetric difference of the advanced and retarded fields is taken, which of course seemingly goes against causality as the advanced field acts from the future backwards in time. It must be mentioned that Dirac's prescription lead to the so called runaway solutions, with the electron acquiring larger and larger velocities in the absense of an external force. This he related to the infinite self energy of the point electron.\\
As far as the breakdown of causality is concerned, this takes place within a period $\sim \tau$, the Compton time. It was at this stage that Wheeler and Feynman reformulated the above action at a distance formalism in terms of what has been called their Absorber Theory. In their formulation, the field that a charge would experience because of its action at a distance on the other charges of the universe, which in turn would act back on the original charge is given by
\begin{equation}
R_e = \frac{2e^2}{3} \vec{x}\label{e4}
\end{equation}
The interesting point is that instead of considering the above force in (\ref{e4}) at the charge $e$, if we consider the responses in its neighbourhood, in fact a neighbourhood at the Compton scale, as was argued recently \cite{iaad}, the field would be precisely the Dirac field given in (\ref{e2}) and (\ref{e3}). The net force emanating from the charge is thus given by
\begin{equation}
F^{\mbox{ret}} = \frac{1}{2} \left\{ F^{\mbox{ret}} + F^{\mbox{adv}}\right\} + \frac{1}{2} \left\{F^{\mbox{ret}} - F^{\mbox{adv}}\right\}\label{e5}
\end{equation}
which is the causal acceptable retarded field. The causal field now consists of the time symmetric field of the charge $e$ together with the Dirac field, that is the second term in (\ref{e5}), which represents the response of the rest of the charges. Interestingly in this formulation we have used a time symmetric field, viz., the first term of (\ref{e5}) to recover the retarded field with the correct arrow of time.\\
There are two important inputs which we can see in the above formulation. The first is the action of the rest of the universe at a given charge and the other is spacetime intervals which are of the order of the Compton scale. Infact we can push the above calculations further. The work done on a charge $e$ at $O$ by the charge at $P$ a distance $r$ in causing a displacement $x$ is given by
\begin{equation}
\frac{e^2x}{r^2} dx\label{e6}
\end{equation}
Now the number of particles at distance $r$ from $O$ is given by
\begin{equation}
n(r) = \rho(r) \cdot 4\pi^2 drUcrcR\label{e7}
\end{equation}
where $\rho(r)$ is the density of particles. So using (\ref{e7}) in (\ref{e6}) the total work is given by
\begin{equation}
E = \int \int \frac{e^2}{r^2} cr \rho 4\pi^2 dr\label{e8}
\end{equation}
which can be shown to be $\sim mc^2$. We thus recover in (\ref{e8}) the inertial energy of the particle in terms of its electromagnetic interactions with the rest of the universe in an action at a distance scheme. Interestingly this can also be deduced in the context of gravitation: The work done on a particle of mass $m$ which we take to be a pion, a typical elementary particle, by the rest of the particles (pions) in the universe is given by
\begin{equation}
\frac{Gm^2N}{R}\label{e9}
\end{equation}
It is known that in (\ref{e9}) $N \sim 10^{80}$ while $R \sim \sqrt{N}l$, the well known Weyl-Eddington formula. Whence the gravitational energy of the pion is given by
\begin{equation}
\frac{Gm^2\sqrt{N}}{l} = \frac{e^2}{l} \sim mc^2\label{e10}
\end{equation}
where in (\ref{e10}) we have used the fact that
\begin{equation}
Gm^2 \sim \frac{e^2}{\sqrt{N}}\label{e11}
\end{equation}
(It must be mentioned that though the Eddington formula and (\ref{e11}) were empirical, they can infact be deduced from theory \cite{cu}, as we will see shortly.) 
\section{The Machian Universe}
This dependence of the mass of a particle on the rest of the universe was argued by Mach in the nineteenth century itself in what is now famous as Mach's Principle \cite{mwt,jv}. The Principle is counter intuitive in that we consider the mass which represents the quantity of matter in a particle to be an intrisic property of the particle. But the following statement of Mach's Principle shows it to be otherwise.\\
If there were no other particles in the universe, then the force acting on the particle $P$ would vanish and so we would have by Newton's second law
\begin{equation}
m\vec{a} = O\label{e12}
\end{equation}
Can we conclude that the acceleration of the particle vanishes? Not if we do not postulate the existence of an absolute background frame in space. In the absense of such a Newtonian absolute space frame, the acceleration $\vec{a}$ would infact be arbitrary, because we could measure this acceleration with respect to arbitrary frames. Then (\ref{e12}) implies that $m = 0$. That is, in the absense of any other matter in the universe, the mass of a material particle would vanish. From this point of view the mass of a particle depends on the rest of the material content of the universe. This has been brought out by the above calculations in (\ref{e8}) and (\ref{e10}).\\
Though Einstein was an admirer of Mach's ideas, his Special Theory of Relativity went counter to them. He subscribed to the concept of Locality according to which information about a part of the universe can be obtained by dealing with that part without taking into consideration the rest of the universe at the same time. In his words, \cite{singh} ``But one one supposition we should, in my opinion absolutely hold fast: the real factual situation of the system $S_2$ is independent of what is done with the system $S_1$ which is spatially separated from the former.''Further, Causality is another cornerstone in Einstein's Physics.
\section{The Quantum Universe}
The advent of Quantum Mechanics however threw up several counter intuitive ideas and Einstein could not reconcile to them. One of these ideas was the wave particle duality. Another was that of the collapse of the wave function in which process Causality becomes a casuality. To put it simply, if the wave function is a super position of the eigen states of an observable then a measurement of the observable yields one of the eigen values no doubt, but it is not possible to predict which one. Due to the observation, the wave function instantly collapses to any one of its eigen states in an acausal manner. To put it another way, the wave function obeys the causal Schrodinger equation, for example, till the instant of observation at which point, causality ceases.\\
Another important counter intuitive feature of Quantum Mechanics is that of non locality. In fact Einstein with Podolsky and Rosen put forward in 1935 his arguments for the incompleteness of Quantum Mechanics on this score \cite{singh,EPR}. This has later come to be known as the EPR paradox. To put it in a simple way, without sacrificing the essential concepts, let us consider two elementary particles, for example two protons kept together somehow. They are then released and move in opposite directions. When the first proton reaches the point $A$ its momentum is measured and turns out to be say, $\vec{p}$. At that instant we can immediately conclude, without any further measurement that the momentum of the second proton which is at the point $B$ is $-\vec{p}$. This follows from the Conservation of Linear Momentum, and is perfectly acceptable in Classical Physics, in which the particles possess a definite momentum at each instant.\\
In Quantum Physics, the difficulty is that we cannot know the momentum at $B$ until and after a measurement is actually performed, and then that value of the momentum is unpredictable. What the above experiment demonstrates is that the proton at $B$ instantly came to have the value $-\vec{p}$ for its momentum when the momentum of the proton at $A$ was measured. This ``instant'' or ``spooky action at a distance'' feature was unacceptable to Einstein.\\
In Quantum Theory however this is legitimate because of another counter intuitive feature which is called Quantum Nonseparability. That is if two systems interact and then separate to a distance, they still have a common state vector. This goes against the concept of Locality and Causality, because it implies instantaneous interaction between distant systems. So in the above example, even though the protons at $A$ and $B$ may be separated, they still have a common wave function which collapses with the measurement of the momentum of any one of them and selfconsistently provides an explanation. This Nonseparability has been characterised by Schrodinger in the following way: ``I would not call that \underline{one}, but rather \underline{the} characteristic of Quantum Mechanics.'' For Einstein however this was like spooky action at a distance. All this has been experimentally verified since 1980 which sets at rest Einstein's objections.\\
However this ``entanglement'' as it is called these days, between distant objects in the universe, does not really manifest itself. An explanation for this was given by Schrodinger himself who argued in effect that entanglement is perfectly legitimate and observable in a universe that consists of let us say just two particles. But a measurement destroys the entanglement. Now in the universe as there are so many particles and correspondingly a huge amount of interference, the entanglement is considerably weakened. What is these days called decoherence works along these lines. This is infact the explanation of the famous ``Schrodinger's Cat'' paradox.\\
This paradox can be explained in the following simple terms: A cat is in an enclosure along with, let us say a microscopic amount of radioactive material. If this material decays, emitting let us say an electron, the electron would fall on a vial of cyanide, releasing it and killing the cat in the process. Let us say that there is a certain probability of such an electron being emitted. So there is the same probability for the cat to be killed. There is also a probability that the electron is not emitted, so that there is the same probability for the cat to remain alive. The cat is therefore in a state which is a superposition of the alive and dead states. It is only when an observer makes an observation that this superposed wave function collapses into either the dead cat state or the alive and kicking cat state, and this happening is acausal. So it is only on an observation being made that the cat is killed or saved, and that too in an unpredictable manner. Till the observation is made the cate is described by the superposed wave function and is thus neither alive nor dead.\\
The resolution of this paradox is of course quite simple. The paradox is valid if the system consists of such few particles and at such distances that they do not interact with each other. Clearly in the real world this idealization is not possible. There are far too many particles and interferences taking place all the time and the superposed wave function would have collapsed almost instantly. This role of the environment has come to be called decoherence. We will return to this point shortly.\\
The important point is that all of Classical and Quantum Physics is based on such idealized laws as if there were no interferences present, that is what may be called a two body scenario is implicit. Clearly this is not a real life scenario.
\section{The Zero Point Field}
Another counter intuitive concept which Quantum Theory introduces is that of the Zero Point Field or Quantum Vacuum. If there were a vacuum, in which at a given point the momentum (and energy) would vanish, then by the Heisenberg Principle, the point itself becomes indeterminate. More realistically, in the vacuum the average energy vanishes but there are fluctuations-- these are the Zero Point Fluctuations. A more classical way of looking at this is that the source free vacuum electromagnetic equations have non zero solutions, in addition to the zero solutions. Interestingly we can argue that the Zero Point Field leads to a minimum interval at the Compton scale \cite{def}.\\
The manifestation of the Zero Point Field has been experimentally tested in what is called the Lamb Shift, which is caused by the fact that the Zero Point Field buffets an ordinary electron in an atom. It has also been verified in the famous Casimir effect \cite{mes,mdef}. The Zero Point Field in this case manifests itself as an attractive force between two parallel plates.\\
Interestingly based on such a Quantum Vacuum and the minimum spacetime intervals the author had proposed a cosmological model in 1997 which predicted an accelerating universe and  a small cosmological constant. In addition, several so called large number relations which had been written off as inexplicable empirical coincidences were shown to follow from the theory \cite{ijmpa}. At that time the prevailing cosmological model was one of dark matter and a decelerating universe. Observational confirmation started coming for the new predictions from 1998 itself while the observational discovery of dark energy, which displaces dark matter, was the scientific Breakthrough of the Year 2003 of the American Association for Advancement of Science \cite{science}.\\
It may be observed that the idea of the Zero Point Field was introduced as early as in 1911 by Max Planck himself to which he assigned an energy $\frac{1}{2} \hbar \omega$. Nernst, a few years later extended these considerations to fields and believed that there would be several interesting consequences in Thermodynamics and even Cosmology.\\
Infact later authors argued that there must be fluctuations of the Quantum Electromagnetic Flield, as required by the Heisenberg Principle, so that we have for an extent $\sim L$
\begin{equation}  
(\Delta B)^2 \geq \hbar c/L^4\label{e13}
\end{equation}
Whence from (\ref{e13}), the dispersion in energy in the entire volume $\sim L^3$ is given by
\begin{equation}
\Delta E \sim \hbar c/L\label{e14}
\end{equation}
(It should be noticed that if $L$ is the Compton wavelength, then (\ref{e14}) gives us the energy of the particle.)Interestingly Braffort and coworkers deduced the Zero Point Field from the Absorber Theory of Wheeler and Feynman, which we encountered earlier. In the process they found that the spectral density of the vacuum field was given by \cite{depena}
\begin{equation}
\rho (\omega) = \mbox{const}\cdot \omega^3\label{e15}
\end{equation}
There have been other points of view which converge to the above conclusions. In any case as we have seen a little earlier, it would be too much of an idealization to consider an atom or a charged particle to be an isolated system. It is interacting with the rest of the universe and this produces a random field.\\
It has also been shown that the constant of proportionality in (\ref{e15}) is given by (Cf.ref.\cite{depena})
$$\frac{\hbar}{2\pi^2 c^3}$$
Interestingly such a constant is implied by Lorentz invariance.\\
From the point of view of Quantum Electrodynamics we reach conclusions similar to those seen above. As Feynman and Hibbs put it \cite{feynman} ``Since most of the space is a vacuum, any effect of the vacuum-state energy of the electromagnetic field would be large. We can estimate its magnitude. First, it should be pointed out that some other infinities occuring in quantum-electrodynamic problems are avoided by a particular assumption called the \underline{cutoff rule}. This rule states that those modes having very high frequencies (short wavelength) are to be excluded from consideration. The rule is justified on the ground that we have no evidence that the laws of electrodynamics are obeyed for wavelengths shorter than any which have yet been observed. In fact, there is a good reason to believe that the laws cannot be extended to the short-wavelength region.\\
``Mathematical representations which work quite well at longer wavelengths lead to divergences if extended into the short wavelength region. The wavelengths in question are of the order of the Compton wavelength of the proton; $1/2\pi$ times this wavelength is $\hbar/mc \simeq 2 \times 10^{-14}cm$.\\
``For our present estimate suppose we carry out sums over wave numbers only up to the limiting value $k_{max} = mc/\hbar$. Approximating the sum over states by an integral, we have, for the vacuum-state energy per unit volume,
$$\frac{E_e}{\mbox{unit \, vol}} = 2 \frac{\hbar c}{2(2\pi)^3} \int^{k_{max}}_0 k 4\pi k^2 dk - \frac{\hbar c k^4_{max}}{8\pi^2}$$
``(Note the first factor $2$, for there are two modes for each $k$). The equivalent mass of this energy is obtained by dividing the result by $c^2$. This gives
$$\frac{m_0}{\mbox{unit \, vol}} = 2 \times 10^{15} g/cm^3$$
Such a mass density would, at first sight at least, be expected to produce very large gravitational effects which are not observed. It is possible that we are calculating in a naive manner, and, if all of the consequences of the general theory of relativity (such as the gravitational effects produced by the large stresses implied here) were included, the effects might cancel out; but nobody has worked all this out. It is possible that some cutoff procedure that not only yields a finite energy density for the vacuum state but also provides relativistic invariance may be found. The implications of such a result are at present completely unknown.\\
``For the present we are safe in assigning the value zero for the vacuum-state energy density. Up to the present time no experiments that would contradict this assumption have been performed.''\\
However the high density encountered above is perfectly meaningful if we consider the Compton scale cut off: Within this volume the density gives us back the mass of an elementary particle like the pion. All this can be put into perspective in the following way. It has been shown in detail by the author that the universe can be considered to have an underpinning of ZPF oscillators at the Planck scale \cite{bgsfpl}. Indeed in all recent approaches towards a unified formulation of gravitation and electromagnetism (including String Theory), the differentiable spacetime manifold of Classical Physics and Quantum Physics has been abandoned and we consider the minimum Planck scale $\sim 10^{-33}cms$ and $10^{-42}secs$ \cite{uof}. We can then show that the universe is a coherent mode of $\bar{N} \sim 10^{120}$ Planck oscillators, spaced a distance $l_P$ apart, that is at the Planck scale. Then the spatial extent is given by
\begin{equation}
R = \sqrt{\bar{N}}l_P\label{e16}
\end{equation}
The mass of the universe is given by
\begin{equation}
M = \sqrt{\bar{N}} m_P\label{e17}
\end{equation}
where $m_P$ is the Planck mass. Moreover we can show that a typical elementary particle like the pion is the ground state of $n \sim 10^{40}$ oscillators and we have (Cf.ref.\cite{bgsfpl})
\begin{equation}
m = \frac{m_P}{\sqrt{n}}\label{e18}
\end{equation}
\begin{equation}
l = \sqrt{n} l_P\label{e19}
\end{equation}
There are $N \sim 10^{80}$ such elementary particles in the universe. Whence we have
\begin{equation}
M = Nm\label{e20}
\end{equation}
We note that equations like (\ref{e16}) and (\ref{e19}) have the Brownian Random Walk characters. At this stage we see asymmetry between equations (\ref{e17}), (\ref{e18}) and (\ref{e20}). The reason is that the universe is an excited state of $\bar{N}$ oscillators whereas an elementary particle is a stable ground state of $n$ Planck oscillators. Furthermore, let us denote the state of each Planck oscillator by $\phi_n$; then the state of the universe can be described in the spirit of entanglement discussed earlier by
\begin{equation}
\psi = \sum_{n} c_n \phi_n,\label{e21}
\end{equation}
$\phi_n$ can be considered to be eigen states of energy with eigen values $E_n$. It is known that (\ref{e21}) can be written as \cite{bgscsf}
\begin{equation}
\psi = \sum_{n} b_n \bar{\phi}_n\label{e22}
\end{equation}
where $|b_n|^2 = 1 \, \mbox{if}\, E < E_n < E + \Delta$ and $= 0$ otherwise under the assumption
\begin{equation}
\overline{(c_n,c_m)} = 0, n \ne m\label{e23}
\end{equation}
(Infact $n$ in (\ref{e23}) could stand for not a single state but for a set of states $n_\imath$, and so also $m$). Here the bar denotes a time average over a suitable interval. This is the well known Random Phase Axiom and arises due to the total randomness amongst the phases $c_n$. Also the expectation value of any operator $O$ is given by
\begin{equation}
< O > = \sum_{n} |b_n|^2 (\bar{\phi}_n, O \bar{\phi}_n)/\sum_{n} |b_n|^2\label{e24}
\end{equation}
Equations (\ref{e22}) and (\ref{e24}) show that effectively we have incoherent states $\bar{\phi}_1, \bar{\phi}_2,\cdots$ once averages over time intervals for the phases $c_n$ in (\ref{e23}) vanish owing to their relative randomness. In the light of the preceding discussion of random fluctuations, we can interpret all this meaningfully: We can identify $\phi_n$ with the ZPF. The time averages are the same as Dirac's zitterbewegung averages over intervals $\sim \frac{\hbar}{mc^2}$ (Cf.ref.\cite{cu}). We then get disconnected or incoherent particles from a single background of vacuum fluctuations exactly as before. The incoherence arises because of the well known random phase relation (\ref{e23}), that is after averating over the suitable interval. Here the entanglement is weakened by the interactions and hence we have (\ref{e20}) for elementary particles, rather than (\ref{e17}).\\
How do we characterize time in this scheme? To consider this problem, we observe that the ground state of $\bar{N}$ Planck oscillators considered above would be, exactly as in (\ref{e18}),
\begin{equation}
\bar {m} = \frac{m_P}{\sqrt{\bar{N}}} \sim 10^{-65}gms\label{ex2}
\end{equation}
The universe is an excited state and consists of $\bar{N}$ such ground state
levels and so we have, from (\ref{ex2})
$$M = \bar{m} \bar{N} = \sqrt{\bar{N}} m_P \sim 10^{55}gms,$$ as required, $M$
being the mass of the universe. Interestingly, the Compton wavelength and time of $\bar{m}$ turn out to be the radius and age of the universe.\\ 
Due to the fluctuation $\sim
\sqrt{n}$ in the levels of the $n$ oscillators making up an elementary
particle, the energy is, remembering that $mc^2$ is the ground state,
$$\Delta E \sim \sqrt{n} mc^2 = m_P c^2,$$ 
 and so the indeterminacy time is
$$\frac{\hbar}{\Delta E} = \frac{\hbar}{m_Pc^2} = \tau_P,$$ as indeed
we would expect.\\ 
The corresponding minimum indeterminacy length
would therefore be $l_P$.  We thus recover the Planck scale. One of the consequences of the minimum
spacetime cut off is that the Heisenberg Uncertainty
Principle takes an extra term. Thus we have,
\begin{equation}
\Delta x \approx \frac{\hbar}{\Delta p} + \alpha \frac{\Delta
p}{\hbar},\, \alpha = l^2 (\mbox{or}\, l^2_P)\label{ex6}
\end{equation}
where $l$ (or $l_P$) is the minimum interval under consideration (Cf.\cite{cu,uof}). 
The first term gives the usual Heisenberg Uncertainty
Principle.\\
Application of the time analogue of (\ref{ex6}) for the
indeterminacy time $\Delta t$ for the fluctuation in energy $\Delta
\bar{E} = \sqrt{N} mc^2$ in the $N$ particle states of the universe
gives exactly as above,
$$\Delta t = \frac{\Delta E}{\hbar} \tau^2_P =
\frac{\sqrt{N}mc^2}{\hbar} \tau^2_P = \frac{\sqrt{N}
m_Pc^2}{\sqrt{n}\hbar} \tau^2_P = \sqrt{n} \tau_P = \tau ,$$ 
In other words, for the fluctuation
$\sqrt{N}$, the time is $\tau$. It must be re-emphasized that the
Compton time $\tau$ of an elementary particle, is an interval within
which there are unphysical effects like zitterbewegung - as pointed
out by Dirac, it is only on averaging over this interval, that we
return to meaningful Physics.  This gives us,
\begin{equation}
dN/dt = \sqrt{N}/\tau\label{ex3}
\end{equation}
On the other hand
due to the fluctuation in the $\bar{N}$ oscillators
constituting the universe, the fluctuational energy is similarly given
by
$$\sqrt{\bar {N}} \bar {m} c^2,$$ which is the same as (\ref{ex2})
above. Another way of deriving (\ref{ex3}) is to observe that as
$\sqrt{n}$ particles appear fluctuationally in time $\tau_P$ which is,
in the elementary particle time scales, $\sqrt{n} \sqrt{n} = \sqrt{N}$
particles in $\sqrt{n} \tau_P = \tau$. That is, the rate of the
fluctuational appearance of particles is
$$
\left(\frac{\sqrt{n}}{\tau_P}\right) = \frac{\sqrt{N}}{\tau} = dN/dt$$
which is (\ref{ex3}). From here by integration,
$$T = \sqrt{N} \tau$$ $T$ is the time elapsed from $N = 1$ and $\tau$
is the Compton time. This gives $T$ its origin in the fluctuations -
there is no smooth ``background'' (or ``being'') time - the root of
time is in ``becoming''. It is the time of a Brownian Wiener
process: A step $l$ gives a step in time $l/c \equiv \tau$ and
therefore the Brownian relation $\Delta x = \sqrt{N} l$ gives $T = \sqrt{N} \tau$ (Cf.refs.\cite{bgsfpl} and \cite{uof}). Time is
born out of acausal fluctuations which are random and therefore
irreversible. Indeed, there is no background time. Time is
proportional to $\sqrt{N}$, $N$ being the number of particles which
are being created spontaneously from the ZPF by fluctuations to the higher energy states of the coherent $\bar{N}$ Planck oscillators.
\section{The Underpinning of the Universe}
So our description of the universe at the Planck scale is that of an entangled wave function as in (\ref{e21}). However we percieve the universe at the elementary particle or Compton scale, where the random phases would have weakened the entanglement, and we have the description as in (\ref{e22}) or (\ref{e24}). Does this mean that $N$ elementary particles in the universe are totally incoherent in which case we do not have any justification for treating them to be in the same spacetime? We can argue that they still interact amongst each other though in comparison this is ``weak''. For instance let us consider the background ZPF whose spectral frequency is given by (\ref{e15}). If there are two particles at $A$ and $B$ separated by a distance $r$, then those wavelengths of the ZPF which are atleast $\sim r$ would connect or link the two particles. Whence the force of interaction between the two particles is given by, remembering that $\omega \propto \frac{1}{r}$,
\begin{equation}
\mbox{Force}\, \quad  \propto \int^\infty_r \omega^3 dr \propto \frac{1}{r^2}\label{e27}
\end{equation}
Thus from (\ref{e27}) we are able to recover the familiar Coulomb Law of interaction. The background ZPF thus enables us to recover the action at a distance formulation. Infact a similar argument can be given \cite{fisch} to recover from QED the Coulomb Law--here the carriers of the force are the virtual photons, that is photons whose life time is within the Compton time of uncertainty permitted by the Heisenberg Uncertainty Principle.\\
It is thus possible to synthesize the field and action at a distance concepts, once it is recognized that there are minimum spacetime intervals at the Compton scale \cite{iaad}. Many of the supposed contradictions arise because of our characterization in terms of spacetime points and consequently a differentiable manifold. Once the minimum cut off at the Planck scale is introduced, this leads to the physical Compton scale and a unified formulation free of divergence problems. We now make a few comments.\\
We had seen that the Dirac formulation of Classical Electrodynamics needed to introduce the acausal advanced field in (\ref{e3}). However the acausality was again within the Compton time scale. Infact this fuzzy spacetime can be modelled by a Wiener process as discussed in \cite{uof}(Cf. also \cite{nottale}). The point here is that the backward and forward time derivatives for $\Delta t \to 0^-$ and $0^+$ respectively do not cancel, as they should not, if time is fuzzy. So we automatically recover from the electromagnetic potential the retarded field for forward derivatives and the advanced fields for backward derivatives. In this case we have to consider both these fields. Causality however is recovered as in (\ref{e5}). This is a transition to intervals which are greater in magnitude compared to the Compton scale.\\
It must also be mentioned that a few assumptions are implicit in the conventional theory using differentiable spacetime manifolds. In the variational problem we use the conventional $\delta$ (variation) which commutes with the time derivatives. So such an operator is constant in time. So also the energy momentum operators in Dirac's displacement operators theory are the usual time and space derivatives of Quantum Theory. But here the displacements are ``instantaneous''. They are valid in a stationary or constant energy scenario, and it is only then that the space and time operators are on the same footing as required by Special Relativity \cite{davydov}. Infact it can be argued that in this theory we neglect intervals $\sim 0(\delta x^2)$ but if $\delta x$ is of the order of the Compton scale and we do not neglect the square of this scale, then the space and momentum coordinates become complex indicative of a noncommutative geometry which has been discussed in detail \cite{bgscst,bgsknmetric,uof}. What all this means is that it is only on neglecting $0(l^2)$ that we have the conventional spacetime of Quantum Theory, including relativistic Quantum Mechanics and Special Relativity, that is the Minkowski spacetime. Coming to the conservation laws of energy and momentum these are based on translation symmetries \cite{roman}-- what it means is the operators $\frac{d}{dx} \, \mbox{or}\,  \frac{d}{dt}$ are independent of $x$ and $t$. There is here a homogeneity property of spacetime which makes these laws non local. This has to be borne in mind, particularly when we try to explain the EPR paradox.\\
The question how a ``coherent'' spacetime can be extracted out of the particles of the universe could be given a mathematical description along the following lines: Let us say that two particles $A$ and $B$ are in a neighbourhood, if they interact at any time. We also define a neighbourhood of a point or particle $A$ as a subset of all points or particles which contains $A$ and at least one other point. If a particle $C$ interacts with $B$ that is, is in a neighbourhood of $B$, then we would say that it is also in the neighbourhood of $A$. That is we define the transitivity property for neighbourhoods. We can then assume the following property \cite{bgsaltaisky}:\\
Given two distinct elements (or even subsets) $A$ and $B$, there is a neighbourhood $N_{A_1}$ such that $A$ belongs to $N_{A_1}$, $B$ does not belong to $N_{A_1}$ and also given any $N_{A_1}$, there exists a neithbourhood $N_{A_\frac{1}{2}}$ such that $A \subset N_{A_\frac{1}{2}} \subset N_{A_1}$, that is there exists an infinite sequence of neithbourhoods between $A$ and $B$. In other words we introduce topological ``closeness''. Alternatively, we could introduce the reasonable supposition that these are a set of Borel subsets.\\
From here, as in the derivation of Urysohn's lemma \cite{simmons}, we could define a mapping $f$ such that $f(A) = 0$ and $f(B) = 1$ and which takes on all intermediate values. We could now define a metric, $d(A,B) = |f(A) - f(B)|$. We could easily verify that this satisfies the properties of a metric.\\
It must be remarked that the metric turns out to be again, a result of a global or a series of large sets, unlike the usual local picture in which is is the other way round.    

\end{document}